\documentclass[10pt,conference]{IEEEtran}

\usepackage{cite}
\usepackage{amsmath,amssymb,amsfonts,amsthm}
\usepackage{algorithm}
\usepackage{algpseudocode}
\usepackage{graphicx}
\usepackage{textcomp}
\usepackage{xcolor}
\usepackage{cleveref}
\usepackage{nccmath}
\usepackage{enumitem}
\usepackage{flushend}
\usepackage{tikz}

\newtheorem{theorem}{Theorem}

\newtheorem{definition}{Definition}

\newtheorem{assumption}{Assumption}

\newtheorem{corollary}{Corollary}

\newtheorem{lemma}{Lemma}

\newtheorem{proposition}{Proposition}

\makeatletter
\newcommand\fs@betterruled{
    \def\@fs@cfont{\bfseries}\let\@fs@capt\floatc@ruled
    \def\@fs@pre{\vspace*{5pt}\hrule height 0.7pt depth0pt \kern2pt}
    \def\@fs@post{\kern2pt\hrule\relax}
    \def\@fs@mid{\hrule\kern2pt}
    \let\@fs@iftopcapt\iftrue}
\floatstyle{betterruled}
\restylefloat{algorithm}
\makeatother

\usepackage[font=small]{caption}
\graphicspath{{images/}}

\newcommand{\overbar}[1]{\mkern 1.5mu\overline{\mkern-1.5mu#1\mkern-3mu}\mkern 1.5mu}
\DeclareMathOperator*{\argmin}{arg\,min}
\IEEEoverridecommandlockouts
\IEEEpubid{\begin{tikzpicture}[remember picture,overlay]
\node[draw=red, rectangle, align=center, text width=\textwidth, text=red, anchor=north] at ([yshift=-10pt]current page.north) {
\small 
© 2025 IEEE. Published in the proceedings of the 2025 IEEE Global Communications Conference (GLOBECOM 2025). The final authenticated version is available online at: https://doi.org/10.1109/GLOBECOM59602.2025.11431993
};
\end{tikzpicture}}

\begin{document}

\title{Joint Max-Min Power Control and Clustering in Cell-Free Wireless Networks: Design and Analysis}

\author{
\IEEEauthorblockN{Achini Jayawardane, Rajitha Senanayake, Erfan Khordad and Jamie Evans}
\IEEEauthorblockA{Department of Electrical and Electronic Engineering, University of Melbourne, Australia\\
Email: awijenayakaj@student.unimelb.edu.au, \{rajitha.senanayake, erfan.khordad, jse\}@unimelb.edu.au}
}
\maketitle
\IEEEpubidadjcol

\begin{abstract}
Cell-free wireless networks have attracted significant interest for their ability to eliminate cell-edge effects and deliver uniformly high service quality through macro-diversity. In this paper, we develop an algorithm to jointly optimize uplink transmit powers and dynamic user-centric access point (AP) clusters in a centralized cell-free network. This approach aims to efficiently mitigate inter-user interference and achieve higher max-min signal-to-interference-plus-noise ratio (SINR) targets for users. To this end, we re-purpose an iterative power control algorithm based on non-linear Perron-Frobenius theory and prove its convergence for the maximum ratio combiner (MRC) receiver under various AP subset selection schemes. We further provide analytical results by framing the joint optimization as a conditional eigenvalue problem with power and AP association constraints, and leveraging Perron-Frobenius theory on a centrally constructed matrix. The numerical results highlight that optimizing each user's serving AP cluster is essential to achieving higher max-min SINR targets with the simple MRC receiver.
\end{abstract}


\section{Introduction} \label{intro}

Traditional cellular networks rely on sparse frequency reuse patterns to manage the limited spectrum. However, to meet the growing demand for high-quality wireless services in the fifth-generation (5G) era, network operators have densely deployed base stations (BSs) with aggressive frequency reuse. This leads to increased inter-cell interference, resulting in coverage issues and performance disparities between cell-center and cell-edge users, even with massive multiple-input multiple-output (MIMO) \cite{demir21}. The \textit{cell-free massive MIMO} concept addresses these issues by using distributed access points (APs) without cell boundaries to jointly serve users \cite{Ngo15, Ngo17}, turning interference into useful signal power \cite{demir21}. Moreover, densely deployed APs also shorten user-AP distances, improving channel conditions for edge users and improving robustness against shadowing through macro-diversity \cite{demir21, Molisch22}.

Canonical cell-free systems assumed that each AP accessed network-wide channel state information (CSI) to serve all users, requiring significant fronthaul signaling and computational resources \cite{Ngo15}. However, full cooperation is often unnecessary due to weak signals from distant users and the challenge of acquiring reliable CSI at far-off APs. Recent work often adopts user-centric clustering, which forms dynamic, overlapping AP subsets tailored to each user's location, channel quality, and service demands \cite{Buzzi20}.

Power control is another key technique for realizing the potential of cell-free massive MIMO. It has been extensively studied over decades to manage interference and mitigate the near-far effect, ensuring reliable service quality across users. Cell-free systems pose unique power control challenges, as multiple APs serve each user with varying channel strengths, and are subject to unrestrained interference, unlike in cellular networks \cite{demir21}. Among various strategies, max-min fairness has received significant attention in the cell-free literature, as it aligns with the core objective of providing uniformly high service quality to all users \cite{Ngo24, Ngo17}. Additionally, the absence of cell-edge users prevents this goal's performance degradation.

Earlier studies analyzed signal-to-interference-plus-noise ratio (SINR) using a linear interference model and a non-negative channel gain matrix. This approach enabled the application of Perron-Frobenius theory to design fixed-point iterative power control algorithms for max-min SINR fairness \cite{Zander92_feb} \cite[Ch.~13]{gantmacher59}. These approaches assumed sufficiently high transmit power to ignore receiver noise. Recently, a concave form of non-linear Perron-Frobenius theory was applied to address the max-min fairness problem in wireless networks. It was shown that a class of such problems can be reformulated as conditional eigenvalue problems with concave self-mappings, which can be efficiently solved using a normalized fixed-point iteration that achieves geometric convergence \cite{Zheng16}. In addition, \cite{Miretti22} leveraged the affine structure of certain noisy-case max-min fairness problems. They derived closed-form solutions for the max-min SINR and user power coefficients through an eigenpair formulation based on Perron-Frobenius theory.

Performance gains in cell-free networks are dependent on efficient spatial resource allocation \cite{Buzzi20,Bjornson13}. Our earlier work showed that integrating user-centric AP clustering with power control in a cell-free system using a maximum ratio combiner (MRC) outperforms both canonical cell-free and traditional cellular networks \cite{Jayawardane25, Jayawardane23}. Specifically, it enabled more users to meet a fixed SINR target with minimal power.

In this paper, we adopt the above joint resource optimization to achieve max-min SINR fairness among users. We propose a novel iterative algorithm that extends the cellular power control framework in \cite{Zheng16} to support dynamic user-centric clustering in cell-free networks. We prove its convergence for various clustered MRC receivers at the central processing unit (CPU). We note that, to ensure scalability, cell-free massive MIMO systems often employ low-complexity receivers such as the MRC \cite{Ngo17,demir21}. We also provide analytical results by formulating the joint optimization as a conditional eigenvalue problem with power and AP association constraints, and using Perron-Frobenius theory on a centrally constructed matrix.

\section{System Model} \label{sys_model}

We consider the uplink of a cell-free network where $R$ APs, each with $K$ antennas, jointly serve $N$ single-antenna users over the same time-frequency resource unit. The APs are connected to a CPU via ideal links for joint user symbol detection \cite{demir21}. The received signal $\mathbf{y} \in \mathbb{C}^{\scriptscriptstyle RK}$ at the CPU is
\begin{equation} \label{eq_5_1}
    \mathbf{y} = \sum \limits_{i = 1}^{\scriptscriptstyle N} \sqrt{p_i}\,\mathbf{h}_i s_i + \mathbf{n}.
\end{equation}
Here, $s_i \sim \mathcal{N}_{\mathbb{C}}(0,1)$ is the data symbol from user $i$ with transmitted power $\mathnormal{p_i}$, and $\mathbf{n}\sim \mathcal{N}_{\mathbb{C}}(\mathbf{0},\sigma^2\mathbf{I})$ is the additive white Gaussian noise vector at the $RK$ AP receiver antennas. The channel vector $\mathbf{h}_{i} = [h_{11i},h_{12i},...,h_{\scriptscriptstyle RK \scriptstyle i}]^{\scriptscriptstyle \mathrm{T}} \in \mathbb{C}^{\scriptscriptstyle RK}$ consists of independent Rayleigh fading gains $h_{rki} \sim \mathcal{N}_{\mathbb{C}}(0, g_{rki}^2)$ from user $i$ to all AP antennas, where $g_{rki}$ represents the large-scale fading \cite{Jayawardane25}. Assuming APs are relatively far from users, we set $g_{rki} = g_{ri}, \forall k$ AP antennas \cite{demir21}. In the standard block fading model, $h_{rki}$ remains constant within a coherence block. Cell-free networks with dense AP deployments benefit from long coherence times due to shorter communication ranges \cite[Ch.~2]{demir21}.

Each user $n$ is served by a subset of APs $\mathcal{S}_n \subset \{1,...,R\}$ to reduce fronthaul and computational demands. The CPU employs MRC, using the receive combining vectors $\mathbf{h}_{rn}$ on uplink data signals from APs $r \in \mathcal{S}_n$ to estimate the symbol $s_n$. For convenience, we define diagonal matrices $\mathbf{D}_{rn} \in \{0,1\}^{\scriptscriptstyle K \times K }$, indicating AP-user associations \cite[Ch.~2]{demir21},
\begin{equation} \label{eq_2}
    \mathbf{D}_{rn} = 
    \begin{cases}
    \mathbf{I}_{\scriptscriptstyle K} \quad r \in \mathcal{S}_n \\ 
    \mathbf{0}_{\scriptscriptstyle K} \quad r \notin \mathcal{S}_n.
    \end{cases}
\end{equation}
We define the AP cluster indicator matrix $\mathbf{D}_{n} = \mathrm{diag} \left(\mathbf{D}_{1n},...,\mathbf{D}_{\scriptscriptstyle R \scriptstyle n}\right) \in \{0,1\}^{\scriptscriptstyle RK \times RK }$, indicating the APs serving user $n$. The effective central MRC receiver is $\mathbf{D}_{n} \mathbf{h}_{n}$, and the estimated symbol at the CPU is $\hat{s}_n = \mathbf{h}_{n}^{\scriptscriptstyle \mathrm{H} } \mathbf{D}_{n} \mathbf{y}$. The instantaneous SINR at the CPU for the user $n$ is
\begin{equation} \label{eq_5_2}
    \mathrm{SINR}_n
    = \frac{p_n \left|\mathbf{h}_n^{\scriptscriptstyle \mathrm{H} } \mathbf{D}_n \mathbf{h}_n\right|^2}{\sum \limits_{\substack{i=1 \\ i \neq n}}^{\scriptscriptstyle N} p_i \left| \mathbf{h}^{\scriptscriptstyle \mathrm{H} }_n \mathbf{D}_n \mathbf{h}_i \right |^2 + \sigma^2 \left( \mathbf{h}_n^{\scriptscriptstyle \mathrm{H} } \mathbf{D}_n \mathbf{h}_n\right)}.
\end{equation}

\section{Problem Formulation}

Our goal is to optimize the uplink transmit power $p^*_n$ and the serving AP subset $\mathcal{S}_n$ for each user $n$ to maximize the minimum SINR across all users. The optimal AP cluster $\mathcal{S}_n$ is selected from $\mathcal{D}$, a set of chosen AP cluster indicator matrices that define different AP subset patterns within $\{1,...,R\}$. The max-min SINR fairness problem is formulated as
\begin{equation}
\begin{aligned} \label{eq_5_4}
    \max \limits_{\mathbf{p} \, \in \, \mathbb{R}^{\scriptscriptstyle N}, \, \mathbf{D}_{n} \in \mathcal{D}} &\min \limits_{n} \, \, \mathrm{SINR}_n(\mathbf{p},\mathbf{D}_n) \\
    \mathrm{s.t} \, \, 0 \leq p_n &\leq \, \mathrm{p_{max}}, \, n = 1,...,N,
\end{aligned}
\end{equation}
where $\mathbf{p}$ the vector of user power coefficients. Introducing an auxiliary variable $\gamma$ as the minimum SINR among all users, the problem is reformulated in epigraph form \cite[Sec.~3.1.7]{boyd04},
\begin{equation}
\begin{aligned} \label{eq_5_5}
    \max \limits_{\mathbf{p} \, \in \, \mathbb{R}^{\scriptscriptstyle N}, \, \gamma \geq 0} &\gamma \\
    \mathrm{s.t} \, \, \max \limits_{\mathbf{D}_{n} \in \mathcal{D}} \, \, &\mathrm{SINR}_n(\mathbf{p},\mathbf{D}_n) \geq \gamma, \\
    \, \, 0 \leq p_n \leq \, &\mathrm{p_{max}}, \, n = 1,...,N.
\end{aligned}
\end{equation}

\section{Access Point Subset Selection Schemes} \label{AP_selection_schemes}

This section presents three AP subset selection schemes to determine clustering requirements in a cell-free massive MIMO system with a centralized MRC receiver.

To determine the optimal AP cluster $\mathcal{S}_n$ for user $n$, we limit the search to a candidate set $\mathcal{M}_n \subset \{1,...,R\}$, selected based on large-scale fading. This reduces the complexity of AP subset selection, combining, fronthaul signaling and channel estimation. We use the AP cluster indicator matrix $\mathbf{D}_{n} \in \mathcal{D}_{n}$ to implement various subset patterns within $\mathcal{M}_n$, where $\mathcal{D}_{n}$ is user-specific due to the user-centric nature of $\mathcal{M}_n$.

The \textit{fixed} scheme is the simplest AP subset selection method, where each user $n$ is served by a fixed cluster of all $\left|\mathcal{M}_n\right|$ APs. For user $n$, this scheme is represented by the set $\mathcal{D}^{\scriptscriptstyle \mathrm{F}}_n$, which contains a single matrix $\mathbf{D}_n$. This diagonal matrix contains block $\mathbf{I}_{\scriptscriptstyle K}$ at indices corresponding to $r \in \mathcal{M}_n$ and block $\mathbf{0}_{\scriptscriptstyle K}$ otherwise.

Some suboptimal receivers cannot fully mitigate interference using the available macro-diversity of a cell-free network and their signal processing techniques. To address this, we propose variable cluster size schemes, allowing users to select APs based on their requirements, interference, and propagation conditions. This approach efficiently utilizes the spatial degrees of freedom in a cell-free network to enhance user performance.

The second scheme, the \textit{add AP} scheme, selects the best AP based on large-scale fading (or average signal power) to form the initial cluster. Then, the two best APs are selected to form the next cluster, and following this process, $\left|\mathcal{M}_n\right|$ AP subsets are formed for user $n$. This scheme is represented by the set $\mathcal{D}^{\scriptscriptstyle \mathrm{Ad}}_n$ with $\left|\mathcal{M}_n\right|$ elements, where $\mathbf{D}_n \in \mathcal{D}^{\scriptscriptstyle \mathrm{Ad}}_n$ follows the above pattern with $\mathbf{I}_{\scriptscriptstyle K}$ for selected APs and $\mathbf{0}_{\scriptscriptstyle K}$ otherwise, at the corresponding diagonal indices.

The third scheme, the \textit{exhaustive search} scheme, is a more flexible but computationally intensive variable cluster size approach. It performs an exhaustive search on $\mathcal{M}_n$, and the set $\mathcal{D}^{\scriptscriptstyle \mathrm{Ex}}_n$ contains $2^{\scriptscriptstyle \left|\mathcal{M}_n\right|} - 1$ matrices. Each matrix $\mathbf{D}_n \in \mathcal{D}^{\scriptscriptstyle \mathrm{Ex}}_n$ has binary block patterns $\left(\{\mathbf{0}_{\scriptscriptstyle K},\mathbf{I}_{\scriptscriptstyle K}\}\right)$ at diagonal positions for $r \in \mathcal{M}_n$, except $\mathbf{D}_n = \mathbf{0}_{\scriptscriptstyle RK }$. For $r \notin \mathcal{M}_n$, the diagonal indices are set to $\mathbf{0}_{\scriptscriptstyle K}$.

These schemes are applied to the matrix set $\mathcal{D}$ in \eqref{eq_5_5}, optimizing each per-user AP cluster. We use $\mathcal{D}_n^x$ with the variable $x$ to denote any of the above clustering schemes.

\section{Proposed Algorithm: Power Control and User-Centric Access Point Assignment} \label{power_AP_ass_algo}

We propose a fixed-point algorithm that solves the max-min SINR fairness problem in \eqref{eq_5_5} by jointly optimizing AP clusters and uplink powers in a cell-free system. This approach extends a framework originally designed for max-min fairness power control with fixed BS assignments in cellular networks \cite{Zheng16}.

Under common conditions in practical wireless networks (see Section~\ref{SINR_structure}), the maximum achievable SINR at optimality equals the worst (minimum) SINR among users. Hence, for a given power allocation $\mathbf{p}^* > \mathbf{0}$, let $\gamma^* = \min \limits_{n} \mathrm{SINR}_n^* = \min \limits_{n} \max \limits_{\mathbf{D}_{n} \in \mathcal{D}_n^x} \mathrm{SINR}_n(\mathbf{p}^*,\mathbf{D}_n) > 0$ denote the minimum SINR achieved after optimizing the AP clusters for each user. Additionally, $\gamma^* = \mathrm{SINR}_n^*, \forall n$, with at least one user transmitting at maximum power, is a standard result in max-min SINR optimization \cite{Marzetta16}. The following relationships then hold.
\begin{multline}
    \label{eq_5_8}
    \gamma^* = \min \limits_{n} \max \limits_{\mathbf{D}_{n} \in \mathcal{D}_n^x} \mathrm{SINR}_n(\mathbf{p}^*,\mathbf{D}_n) \\ = \frac{\mathrm{p_{max}}}{\max \limits_{n} \min \limits_{\mathbf{D}_{n} \in \mathcal{D}_n^x} I_n(\mathbf{p}^*,\mathbf{D}_n)} = \frac{\mathrm{p_{max}}}{\max \limits_{n} T_n(\mathbf{p}^*)},
\end{multline}
where
\begin{equation} \label{eq_5_9}
    I_n(\mathbf{p},\mathbf{D}_n) = 
    \frac{\sum \limits_{\substack{i=1 \\ i \neq n}}^{\scriptscriptstyle N} p_i \left| \mathbf{h}^{\scriptscriptstyle \mathrm{H} }_n \mathbf{D}_n \mathbf{h}_i \right |^2 + \sigma^2 \left( \mathbf{h}_n^{\scriptscriptstyle \mathrm{H} } \mathbf{D}_n \mathbf{h}_n\right)}{\left|\mathbf{h}_n^{\scriptscriptstyle \mathrm{H} } \mathbf{D}_n \mathbf{h}_n\right|^2},
\end{equation}
is an interference function and 
\begin{equation} \label{eq_5_10}
    T_n(\mathbf{p}) = \min \limits_{\mathbf{D}_{n} \in \mathcal{D}_n^x} I_n(\mathbf{p},\mathbf{D}_n).
\end{equation}
Since all users attain equal SINR at optimality
\begin{equation} \label{eq_5_11}
    \frac{\mathbf{p}^*}{{\boldsymbol{T}}(\mathbf{p}^*)} = \frac{\mathrm{p_{max}}}{\max \limits_{n} T_n(\mathbf{p}^*)} \mathbf{1},
\end{equation}
where $\frac{\mathbf{p}^*}{\boldsymbol{T}(\mathbf{p}^*)} = \left[\frac{p_1^*}{T_1(\mathbf{p}^*)},...,\frac{p_{\scriptscriptstyle N}^*}{T_{\scriptscriptstyle N}(\mathbf{p}^*)}\right]^{\scriptscriptstyle \mathrm{T}}$ and $\mathbf{1}$ is a vector of all ones. Based on equation~\eqref{eq_5_11}, where $\mathbf{p}^*$ is a solution, the following fixed-point iteration will converge to the optimal solution of problem~\eqref{eq_5_5} by jointly optimizing user transmit powers and AP clusters.
\begin{equation} \label{eq_5_12}
    \mathbf{p}(t) = \frac{\mathrm{p_{max}}}{\max \limits_{n} T_n(\mathbf{p}(t-1))} {\boldsymbol{T}}(\mathbf{p}(t-1)),
\end{equation}
where $t$ is the iteration index. Here, $\frac{\mathrm{p_{max}}}{\max \limits_{n} T_n(\mathbf{p})}$ acts as a scaling factor \cite{Zheng16}, ensuring maximum user power constraints are met.

Algorithm~\ref{5_algo} details our proposed fixed-point algorithm. Iterations continue until relative power updates converge within tolerance $\epsilon$ (step~\ref{line_5_8}). The power update from equation~\eqref{eq_5_12} is performed in step~\ref{line_5_7}, while step~\ref{line_5_5} implements the AP subset selection schemes from Section~\ref{AP_selection_schemes} and computes $T_n(\mathbf{p})$ for each user.

\begin{algorithm}
\caption{
Fixed-Point Algorithm for Joint Uplink Power and User-Centric Cluster Optimization for Max-Min SINR
} \label{5_algo}
\begin{algorithmic} [1]
    \State \textbf{Initialization}:
    Set Initial power vector $\mathbf{p}(0) = \mathbf{p}_0$; Iteration step $t = 0$; Precision $\epsilon$; Set $\mathcal{D}_n^x$ with $x \in \{\mathrm{F, Ad, Ex}\}$;
    \textit{break condition} = \textit{false}.
    \While{\textit{break condition} is \textit{false}}
        \State $t \gets t+1$
        \For {$n=1,...,N$ users}
            \State $T_n(\mathbf{p}(t-1)) \gets \min \limits_{\mathbf{D}_{n} \in \mathcal{D}_n^x} I_n(\mathbf{p}(t-1), \mathbf{D}_n)$ \label{line_5_5}
        \EndFor
        \State $\mathbf{p}(t) \gets \frac{\mathrm{p_{max}}}{\max \limits_{n} T_n(\mathbf{p}(t-1))} {\boldsymbol{T}}(\mathbf{p}(t-1))$ \label{line_5_7}
        \If{$\left|\mathbf{p}(t) - \frac{\mathrm{p_{max}}}{\max \limits_{n} T_n(\mathbf{p}(t))}\boldsymbol{T}(\mathbf{p}(t))\right|/\mathbf{p}(t) <\epsilon$} \label{line_5_8}
            \State \textit{break condition} $\gets$ \textit{true}
        \EndIf
    \EndWhile
\end{algorithmic}
\end{algorithm}

\section{Proof of Convergence}

This section outlines the required SINR properties for applying Algorithm~\ref{5_algo} and then discusses its convergence.

\subsection{Properties of SINR Expressions} \label{SINR_structure}

The user SINRs in \eqref{eq_5_5}, as detailed in \eqref{eq_5_2}, must satisfy the following properties to ensure equivalence between the solutions of \eqref{eq_5_5} and $\mathbf{p}^*$ of \eqref{eq_5_11} \cite[Assumption~1]{Zheng16}.
\begin{assumption} (Competitive Utility Functions): \label{5_assumption}
    \begin{itemize} [itemsep=0pt]
    \item Positivity: For all $n, \max \limits_{\mathbf{D}_{n} \in \mathcal{D}_n^x} \mathrm{SINR}_n(\mathbf{p},\mathbf{D}_n) > 0$ if $\mathbf{p} > \mathbf{0}$ and, $\max \limits_{\mathbf{D}_{n} \in \mathcal{D}_n^x} \mathrm{SINR}_n(\mathbf{p},\mathbf{D}_n) = 0$ if and only if $p_n = 0$.
    \item Competitiveness: For all $n, \max \limits_{\mathbf{D}_{n} \in \mathcal{D}_n^x} \mathrm{SINR}_n(\mathbf{p},\mathbf{D}_n)$ is strictly increasing with respect to $p_n$ and is strictly decreasing with respect to $p_i$, for $i \neq n$, when $p_n > 0$.
    \item Directional Monotonicity: For $\alpha > 1$ and $p_n > 0, \allowbreak \max \limits_{\mathbf{D}_{n} \in \mathcal{D}_n^x} \mathrm{SINR}_n(\alpha \mathbf{p},\mathbf{D}_n) > \max \limits_{\mathbf{D}_{n} \in \mathcal{D}_n^x} \mathrm{SINR}_n(\mathbf{p},\mathbf{D}_n), \forall n$.
    \end{itemize}
\end{assumption}
From \eqref{eq_5_2}, \eqref{eq_5_9}, and \eqref{eq_5_10}, we derive $\max \limits_{\mathbf{D}_{n} \in \mathcal{D}_n^x} \mathrm{SINR}_n(\mathbf{p},\mathbf{D}_n) = \frac{p_n}{T_n(\mathbf{p})}$. The following corollary shows that this SINR expression with optimized AP clusters satisfies Assumption~\ref{5_assumption}.

\begin{corollary} \label{5_corollary}
SINR expressions of the form $\mathrm{SINR}_n(\mathbf{p}) = \frac{p_n}{T_n(\mathbf{p})}$, where $T_n(\mathbf{p})$ is a standard interference function, fulfill the properties in Assumption~\ref{5_assumption}, $\forall \mathbf{p} > \mathbf{0}$.
\end{corollary}

\begin{IEEEproof}
We can first prove the standard interference function properties of our $T_n(\mathbf{p})$ in \eqref{eq_5_10} following \cite[Thm.~1]{Jayawardane23}, then apply \cite[Lemma~1]{Cavalcante23} to verify properties in Assumption~\ref{5_assumption}.
\end{IEEEproof}

\subsection{Convergence of the Algorithm}

Based on non-linear Perron-Frobenius theory \cite{Nussbaum12}, the iteration in \eqref{eq_5_12} is a contraction mapping when the function $\boldsymbol{T}(\mathbf{p})$ is both positive and concave \cite[Thm.~1]{Krause86}. Under these conditions, the global convergence of the iteration in \eqref{eq_5_12} to a unique fixed point is guaranteed \cite{Krause86, Zheng16}. This unique fixed point is the only solution to equation~\eqref{eq_5_11}, which, under Assumption~\ref{5_assumption}, is equivalent to the optimal solution of problem~\eqref{eq_5_5}. However, the concavity of $\boldsymbol{T}(\mathbf{p})$ does not imply that $\max \limits_{\mathbf{D}_{n} \in \mathcal{D}_n^x}{\mathrm{SINR}}_n(\mathbf{p},\mathbf{D}_n)$ is concave for all users $n$, nor does it ensure that problem~\eqref{eq_5_5} is convex \cite{Zheng16}. Since positivity is inherent in standard interference functions \cite{Jayawardane23, Jayawardane25}, we prove the concavity of $\boldsymbol{T}(\mathbf{p}) = \left[{T_1(\mathbf{p})},...,{T_{\scriptscriptstyle N}(\mathbf{p})}\right]^{\scriptscriptstyle \mathrm{T}}$ introduced in \eqref{eq_5_10}.

\begin{theorem}
    $\boldsymbol{T}(\mathbf{p})$ is a concave function.
\end{theorem}
\begin{IEEEproof}
We first write the interference function in \eqref{eq_5_9} as
\begin{equation}
\begin{aligned}
\label{eq_5_15}
    I_n(\mathbf{p},\mathbf{D}_n)
    &= \sum \limits_{\substack{i=1 \\ i \neq n}}^{\scriptscriptstyle N} p_i \left| \frac{  \mathbf{h}^{\scriptscriptstyle \mathrm{H} }_n \mathbf{D}_n \mathbf{h}_i}{\mathbf{h}_n^{\scriptscriptstyle \mathrm{H} } \mathbf{D}_n \mathbf{h}_n}\right|^2+ \frac{\sigma^2}{\left( \mathbf{h}_n^{\scriptscriptstyle \mathrm{H} } \mathbf{D}_n \mathbf{h}_n\right)} \\
    &= \mathbf{z}_n^{\scriptscriptstyle \mathrm{T}}\mathbf{p} + \sigma^2_{n},
\end{aligned}
\end{equation}
where $\mathbf{z}_{n} = [{z}_{n1},...,{z}_{n\scriptscriptstyle N}]^{\scriptscriptstyle \mathrm{T}}$ with elements 
\begin{equation*}
    {z}_{ni} = 
    \begin{cases}
    \left| \frac{\mathbf{h}^{\scriptscriptstyle \mathrm{H} }_n \mathbf{D}_n \mathbf{h}_i}{\mathbf{h}_n^{\scriptscriptstyle \mathrm{H} } \mathbf{D}_n \mathbf{h}_n}\right|^2 & i \neq n \\
    0 & i = n,
\end{cases}
\end{equation*}
and $\sigma^2_n = \frac{\sigma^2}{\left( \mathbf{h}_n^{\scriptscriptstyle \mathrm{H} } \mathbf{D}_n \mathbf{h}_n\right)}$. Since $I_n(\mathbf{p},\mathbf{D}_n)$ is an affine function of $\mathbf{p}$, and $T_n(\mathbf{p})$ is the pointwise minimum of affine functions by \eqref{eq_5_10} and \eqref{eq_5_15}, it follows that $T_n(\mathbf{p})$ is concave \cite{boyd04}. Thus, $T_n(\mathbf{p})$ is concave for all $n$, making $\boldsymbol{T}(\mathbf{p})$ concave.
\end{IEEEproof}

\section{Theoretical Characterization of Max-Min SINR} \label{5_theory_analysis}

In this section, we show that the optimal max-min SINR, power allocation, and AP clusters are characterized by the spectral radius of a centrally constructed matrix.

Rewriting \eqref{eq_5_11} with $\gamma^*$ as the max-min SINR, we obtain
\begin{equation} \label{eq_5_18}
    {\boldsymbol{T}}(\mathbf{p}^*) =  \frac{1}{\gamma^*} \mathbf{p}^*, \quad \gamma^* > 0, \, \mathbf{p}^* > \mathbf{0}, \, \max \limits_{n} {p}_n^* = \mathrm{p_{max}}.
\end{equation}
Assuming a fixed AP assignment for all users, that is ${\boldsymbol{T}}(\mathbf{p}^*) = [{I}_1(\mathbf{p}^*,\mathbf{D}_1),...,{I}_{\scriptscriptstyle N}(\mathbf{p}^*,\mathbf{D}_{\scriptscriptstyle N})]^{\scriptscriptstyle \mathrm{T}}$, we can view \eqref{eq_5_18} as a conditional eigenvalue problem
\begin{equation}
\begin{aligned} \label{eq_5_19}
    \mathrm{Find} \ &\lambda \in \mathbb{R}> 0, \, \mathbf{p} \in \mathbb{R}^{\scriptscriptstyle N} > \mathbf{0} \\
    \mathrm{s.t} \ &{\boldsymbol{T}}(\mathbf{p}) =  \lambda \mathbf{p}, \quad \max \limits_{n} {p}_n = \mathrm{p_{max}}.
\end{aligned}
\end{equation}
For a standard interference mapping ${\boldsymbol{T}}(\mathbf{p})$, the conditional eigenvalue problem has a unique solution \cite[Prop.~1]{Miretti22}.

Equation~\eqref{eq_5_18} and the constraint in \eqref{eq_5_19} reveal an inverse relationship between the optimal max-min SINR ($\gamma^*$) and the eigenvalue solution ($\lambda^*$). By classical Perron-Frobenius theory, the spectral radius $\rho(\mathbf{A})$ of a non-negative matrix $\mathbf{A}$ is the smallest positive eigenvalue with a positive real eigenvector \cite[Ch.~13]{gantmacher59}. Thus, the goal of optimal clustering is to find AP associations that minimize $\rho(\mathbf{Z}_{k})$, subject to the mapping $\boldsymbol{T}(\mathbf{p}) = \mathbf{Z}_{k}\mathbf{p} > 0$, where $\mathbf{Z}_{k}$ is a non-negative matrix.

Let us focus on our $\boldsymbol{T}(\mathbf{p})$. From \eqref{eq_5_15}, we derive $\boldsymbol{I}(\mathbf{p},\mathbf{D}) = [I_1(\mathbf{p},\mathbf{D}_1), ..., I_{\scriptscriptstyle N}(\mathbf{p},\mathbf{D}_{\scriptscriptstyle N})]^{\scriptscriptstyle \mathrm{T}}$, where $\mathbf{D} \in \mathcal{D}^{\scriptscriptstyle N} = \{ \{\mathbf{D}_n\}^{\scriptscriptstyle N}:\mathbf{D}_n \in \mathcal{D}_n^x,\, \forall n \in \{1,...,N\},\,x \in \{\mathrm{F, Ad, Ex}\} \}$, in affine form
\begin{equation} \label{eq_5_20}
    \boldsymbol{I}(\mathbf{p},\mathbf{D}) = \mathbf{Z}\mathbf{p} + \boldsymbol{\sigma},
\end{equation}
where $\mathbf{Z} = [\mathbf{z}_{1}^{\scriptscriptstyle \mathrm{T}},...,\mathbf{z}_{\scriptscriptstyle N}^{\scriptscriptstyle \mathrm{T}}]^{\scriptscriptstyle \mathrm{T}} \in \mathbb{R}^{\scriptscriptstyle N \times \scriptscriptstyle N} \geq \mathbf{0}$ and $\boldsymbol{\sigma} = [\sigma^2_{1},...,\sigma^2_{\scriptscriptstyle N}] \in \mathbb{R}^{\scriptscriptstyle N} > \mathbf{0}$. Then, $\boldsymbol{T}(\mathbf{p}) = \min \limits_{\mathbf{D} \in \mathcal{D}^{\scriptscriptstyle N}} \mathbf{Z}\mathbf{p} + \boldsymbol{\sigma}$. This leads to a new conditional eigenvalue problem
\begin{equation}
\begin{aligned} \label{eq_5_21}
    \mathrm{Find} \ &\lambda \in \mathbb{R}> 0, \, \mathbf{p} \in \mathbb{R}^{\scriptscriptstyle N} > \mathbf{0} \\
    \mathrm{s.t} \ &\min \limits_{\mathbf{D} \in \mathcal{D}^{\scriptscriptstyle N}} \mathbf{Z}\mathbf{p} + \boldsymbol{\sigma} = \lambda \mathbf{p}, \quad \max \limits_{n} {p}_n = \mathrm{p_{max}},
\end{aligned}
\end{equation}
where $\mathbf{Z}$ and $\boldsymbol{\sigma}$ are functions of $\mathbf{D}$ from \eqref{eq_5_15} and \eqref{eq_5_20}. Based on \cite[Prop.~1]{Miretti22}, which addresses a general standard interference mapping, the following proposition holds.

\begin{proposition} \label{proposition_5_1}
    Given the standard interference mapping $\min \limits_{\mathbf{D} \in \mathcal{D}^{\scriptscriptstyle N}} \mathbf{Z}\mathbf{p} + \boldsymbol{\sigma}$, the conditional eigenvalue problem in \eqref{eq_5_21} has a unique solution.
\end{proposition} 

In light of Proposition~\ref{proposition_5_1}, the unique solutions to problem~\eqref{eq_5_5} are characterized in Proposition~\ref{proposition_5_2}. We will use the following known results to prove Proposition~\ref{proposition_5_2}.

\begin{proposition} \label{proposition_5_3}
    Let $\mathbf{A} \in \mathbb{R}^{\scriptscriptstyle N \times N}$ be a non-negative matrix and $\lambda, \mu \in \mathbb{R}> 0, \, \mathbf{x}, \mathbf{y}, \in \mathbb{R}^{\scriptscriptstyle N} > \mathbf{0}$
    \begin{enumerate} [itemsep=0pt]
        \item If $ \exists \mathbf{y}$ s.t $\mathbf{A}\mathbf{y} \leq \lambda \mathbf{y}$, then $\rho(\mathbf{A}) \leq \lambda$
        \cite[Lem.~1.1]{Nussbaum86}.
        \item If $ \exists \mathbf{x}$ s.t $\mu\mathbf{x} \leq \mathbf{A}\mathbf{x}$, then $\mu \leq \rho(\mathbf{A})$ \cite[Thm.~2.4.1]{Krause15}.
    \end{enumerate}
\end{proposition}

\begin{proposition} \label{proposition_5_2}
Denote $\left( \gamma^*, \mathbf{D}^*, \mathbf{p}^* \right)$ as the unique solutions to problem~\eqref{eq_5_5}, which are given by
\begin{equation} \label{eq_5_22}
    \gamma^* = \frac{1}{\min \limits_{\mathbf{D} \in \mathcal{D}^{\scriptscriptstyle N}} \max \limits_{j} \rho\left( \mathbf{Z}_j(\mathbf{D})\right)},
\end{equation}
\begin{equation} \label{eq_5_23}
    \mathbf{D}^* = {\argmin \limits_{\mathbf{D} \in \mathcal{D}^{\scriptscriptstyle N}} \max \limits_{j} \rho\left( \mathbf{Z}_j(\mathbf{D})\right)},
\end{equation}
and
\begin{equation} \label{eq_5_24}
    \mathbf{p}^* = \gamma^* \left(\mathbf{I}_{\scriptscriptstyle RN} - \gamma^* \mathbf{Z}(\mathbf{D}^*) \right)^{-1} \boldsymbol{\sigma} (\mathbf{D}^*),
\end{equation}
where for a given ${\mathbf{D} \in \mathcal{D}^{\scriptscriptstyle N}}$ and $ j \in \{1,...,N\}$,
\begin{equation} \label{eq_5_25}
    \mathbf{Z}_j(\mathbf{D}) = \mathbf{Z}(\mathbf{D}) + \frac{1}{\mathrm{p_{max}}} \boldsymbol{\sigma} (\mathbf{D}) \mathbf{a}_j^{\scriptscriptstyle \mathrm{T}},
\end{equation}
with $\mathbf{a}_j = [a_1,...,a_{\scriptscriptstyle N}]^{\scriptscriptstyle \mathrm{T}}$ s.t $a_i = 0, i \neq j$ and $a_i = 1, i = j$, and with $\mathbf{Z}(\mathbf{D})$ and $\boldsymbol{\sigma} (\mathbf{D})$ indicating their dependence on $\mathbf{D}$.

Moreover, $\mathbf{p}^*$ is the corresponding eigenvector satisfying $\mathbf{p}^* \in \mathbb{R}^{\scriptscriptstyle N} > \mathbf{0}$ and $\max \limits_{n} {p}_n^* = \mathrm{p_{max}}$, associated with the least dominant eigenvalue $\min \limits_{\mathbf{D} \in \mathcal{D}^{\scriptscriptstyle N}} \max \limits_{j} \rho\left( \mathbf{Z}_j(\mathbf{D})\right)$ and its equivalent matrix across various AP associations.
\end{proposition}

\begin{IEEEproof}
The unique solution $\left(\frac{1}{\gamma^*}, \mathbf{D}^*,\mathbf{p}^* \right)$ to the conditional eigenvalue problem~\eqref{eq_5_21} also solves equation~\eqref{eq_5_18}, which is equivalent to solving problem~\eqref{eq_5_5} under Assumption~\ref{5_assumption}. Thus, from \eqref{eq_5_21}, we obtain
\begin{multline} \label{eq_5_26}
    \mathbf{p}^* = \gamma^* \left(\mathbf{Z}(\mathbf{D}^*) \mathbf{p}^* + \boldsymbol{\sigma} (\mathbf{D}^*) \right), \\ \mathbf{p}^* \in \mathbb{R}^{\scriptscriptstyle N} > \mathbf{0}, \max \limits_{n} {p}_n^* = \mathrm{p_{max}}.
\end{multline}
Let us first consider a chosen fixed $\mathbf{D}$. Similarly, we have
\begin{multline} \label{eq_5_266}
    \overbar{\mathbf{p}} = \gamma \left(\mathbf{Z}(\mathbf{D}) \overbar{\mathbf{p}}+ \boldsymbol{\sigma} (\mathbf{D}) \right), \\ \overbar{\mathbf{p}} \in \mathbb{R}^{\scriptscriptstyle N} > \mathbf{0}, \max \limits_{n} \overbar{p}_n = \mathrm{p_{max}}.
\end{multline}
Here, it is possible that $\gamma \leq \gamma^*$ and $\overbar{\mathbf{p}} \neq \mathbf{p}^*$. Now, $\exists k \in \{1,...,N\}$,
\begin{equation} \label{eq_5_27}
    \frac{1}{\mathrm{p_{max}}} \mathbf{a}_j^{\scriptscriptstyle \mathrm{T}}\overbar{\mathbf{p}} \leq \frac{1}{\mathrm{p_{max}}}\mathbf{a}_{k}^{\scriptscriptstyle \mathrm{T}}\overbar{\mathbf{p}} = 1, \quad \forall j \in \{1,...,N\}.
\end{equation}
Following the proof of \cite[Prop.~4]{Miretti22}, $\forall j \in \{1,...,N\}$ we have
\begin{multline}
\label{eq_5_28}
    \mathbf{Z}_j(\mathbf{D})\overbar{\mathbf{p}} = \mathbf{Z}(\mathbf{D})\overbar{\mathbf{p}} + \frac{1}{\mathrm{p_{max}}} \boldsymbol{\sigma} (\mathbf{D}) \mathbf{a}_j^{\scriptscriptstyle \mathrm{T}}\overbar{\mathbf{p}} \\ 
    \leq \mathbf{Z}(\mathbf{D})\overbar{\mathbf{p}} + \frac{1}{\mathrm{p_{max}}} \boldsymbol{\sigma} (\mathbf{D}) \mathbf{a}_{k}^{\scriptscriptstyle \mathrm{T}}\overbar{\mathbf{p}} = \mathbf{Z}_{k}(\mathbf{D})\overbar{\mathbf{p}} = \frac{1}{\gamma} \overbar{\mathbf{p}}.
\end{multline}
This follows first from the inequality in \eqref{eq_5_27}, the positivity of $\boldsymbol{\sigma} (\mathbf{D})$, and then from the equalities in \eqref{eq_5_27} and \eqref{eq_5_266}. Now, from Proposition~\ref{proposition_5_3}\,(1), $\rho\left( \mathbf{Z}_j(\mathbf{D}) \right) \leq \rho\left( \mathbf{Z}_{k}(\mathbf{D}) \right), \, \forall j \in \{1,...,N\}$. We also have
\begin{equation} \label{eq_5_29}
    \mathbf{Z}_{k'}(\mathbf{D})\mathbf{p}^* \geq \min \limits_{\mathbf{D} \in \mathcal{D}^{\scriptscriptstyle N}} \mathbf{Z}_{k'}(\mathbf{D})\mathbf{p}^* = \mathbf{Z}_{k'}(\mathbf{D^*})\mathbf{p}^*,
\end{equation}
where $\frac{1}{\mathrm{p_{max}}} \mathbf{a}_j^{\scriptscriptstyle \mathrm{T}}\mathbf{p}^* \leq \frac{1}{\mathrm{p_{max}}}\mathbf{a}_{k'}^{\scriptscriptstyle \mathrm{T}}\mathbf{p}^* = 1, \, \forall j \in \{1,...,N\}$. 
Here, $k' = k$ or $k' \neq k$. Now, from \eqref{eq_5_26}, we can obtain
\begin{equation} \label{eq_5_30}
    \mathbf{Z}_{k'}(\mathbf{D^*})\mathbf{p}^* 
    = \mathbf{Z}(\mathbf{D}^*) \mathbf{p}^* + \boldsymbol{\sigma} (\mathbf{D}^*)
    = \frac{1}{\gamma^*} \mathbf{p}^*.
\end{equation}
From \eqref{eq_5_30}, it is evident that the solution to problem~\eqref{eq_5_5} corresponds to the eigenpair $(\frac{1}{\gamma^*}, \mathbf{p}^*)$ of the equivalent matrix $\mathbf{Z}_{k'}(\mathbf{D^*})$, satisfying $\frac{1}{\gamma^*} \in \mathbb{R} > 0$ and $\mathbf{p}^* \in \mathbb{R}^{\scriptscriptstyle N} > \mathbf{0}, \max \limits_{n} {p}_n^* = \mathrm{p_{max}}$. Furthermore, from the inequality in \eqref{eq_5_29} and Proposition~\ref{proposition_5_3}\,(2), $\rho\left( \mathbf{Z}_{k'}(\mathbf{D}^*) \right) \leq \rho\left( \mathbf{Z}_{k'}(\mathbf{D}) \right), \, \forall \mathbf{D} \in \mathcal{D}^{\scriptscriptstyle N}$. Also, from our previous deduction, $\rho\left( \mathbf{Z}_{k'}(\mathbf{D}) \right) \leq \rho\left( \mathbf{Z}_{k}(\mathbf{D}) \right)$. Therefore, $\rho\left( \mathbf{Z}_{k'}(\mathbf{D}^*) \right) \leq \rho\left( \mathbf{Z}_{k}(\mathbf{D}) \right)$. Then, from \eqref{eq_5_30},
\begin{equation} \label{eq_5_31}
    \frac{1}{\gamma^*} = \rho\left(\mathbf{Z}_{k'}(\mathbf{D^*})\right) = \min \limits_{\mathbf{D} \in \mathcal{D}^{\scriptscriptstyle N}} \max \limits_{j} \rho\left( \mathbf{Z}_j(\mathbf{D})\right) > 0.
\end{equation}
Thus, \eqref{eq_5_22}, \eqref{eq_5_23}, and the last statement of Proposition~\ref{proposition_5_2} are proved. The power vector in \eqref{eq_5_24} follows from solving equation~\eqref{eq_5_26}. The uniqueness and positivity of this $\mathbf{p}^*$ require $\rho\left( \gamma^*\mathbf{Z}(\mathbf{D}^*) \right) < 1$, ensuring that $\left(\mathbf{I}_{\scriptscriptstyle RN} - \gamma^* \mathbf{Z}(\mathbf{D}^*) \right)$ is invertible with a strictly positive inverse \cite[Thm.~2.1]{Seneta81}. To prove this, we can write from \eqref{eq_5_31},
\begin{equation} \label{eq_5_32}
\frac{\mathbf{Z}(\mathbf{D}^*)}{\rho \left( \mathbf{Z}_{k'}(\mathbf{D}^*)\right)} = \gamma^*\mathbf{Z}(\mathbf{D}^*).
\end{equation}
From \eqref{eq_5_30}, it is clear that $\mathbf{Z}(\mathbf{D}^*)\mathbf{p}^* < \mathbf{Z}_{k'}(\mathbf{D^*})\mathbf{p}^*$ by construction and the positivity of $\boldsymbol{\sigma} (\mathbf{D})$. Hence, from Proposition~\ref{proposition_5_3}\,(1),
$\rho\left( \mathbf{Z}(\mathbf{D}^*) \right) < \rho\left( \mathbf{Z}_{k'}(\mathbf{D}^*) \right)$. Now, from \eqref{eq_5_32}, we conclude $\rho\left( \gamma^*\mathbf{Z}(\mathbf{D}^*) \right) < 1$.
\end{IEEEproof}

The calculation of \eqref{eq_5_22} and \eqref{eq_5_23} is costly due to the many possible combinations of $\mathbf{D}$ for all users. Instead, the proposed iterative algorithm in Section~\ref{power_AP_ass_algo} is preferable.

\section{Numerical Examples} \label{numerical}

This section presents numerical results from the proposed joint optimization algorithm, which aims to achieve max-min SINR across various network setups. The theoretical characterization from Section~\ref{5_theory_analysis} is also validated by comparing it with the algorithm.

The cell-free wireless network consists of APs positioned at fixed locations in a hexagonal grid, with varying numbers of users uniformly distributed across the entire network area. The inter-AP distance is set to 100 m. The channel between user $n$ and antenna $k$ of AP $r$ experiences independent block Rayleigh fading $h_{rkn} \sim \mathcal{N}_{\mathbb{C}}(0, g_{rn}^2)$, with $g_{rn} = -30.5-36.7\,\mathrm{log}_{10} \left( \frac{d_{rn}}{1\,[\mathrm{m}]}\right) + F_{rn} [\mathrm{dB}]$ \cite[Ch.~5]{demir21}. Here, $d_{rn}$ is the 3-dimensional distance from user $n$ to AP $r$, accounting for a 10 m height difference. The path loss exponent is $\alpha = 3.67$, and shadow fading is modeled as $F_{rn} \sim \mathcal{N}(0, 4^2)$, with terms from an AP to different users correlated as 
\begin{equation}
    \mathbb{E}\{F_{rn}F_{sm}\} =
    \begin{cases}
        4^2 2^{-\delta_{nm}/\,9\,[\mathrm{m}]}  &r = s \\
        0 &r \neq s,
    \end{cases}
\end{equation}
where $\delta_{nm}$ is the distance between users $n$ and $m$. The terms are uncorrelated for different APs $r$ and $s$ \cite[Ch.~5]{demir21}. The noise power at each AP antenna is $-94\,\mathrm{dBm}$ \cite[Ch.~5]{demir21}. The numerical analysis uses 5000 uniformly randomized user locations and Rayleigh fading channel realizations.

\begin{figure}
    \centering
    \includegraphics[width=\linewidth]{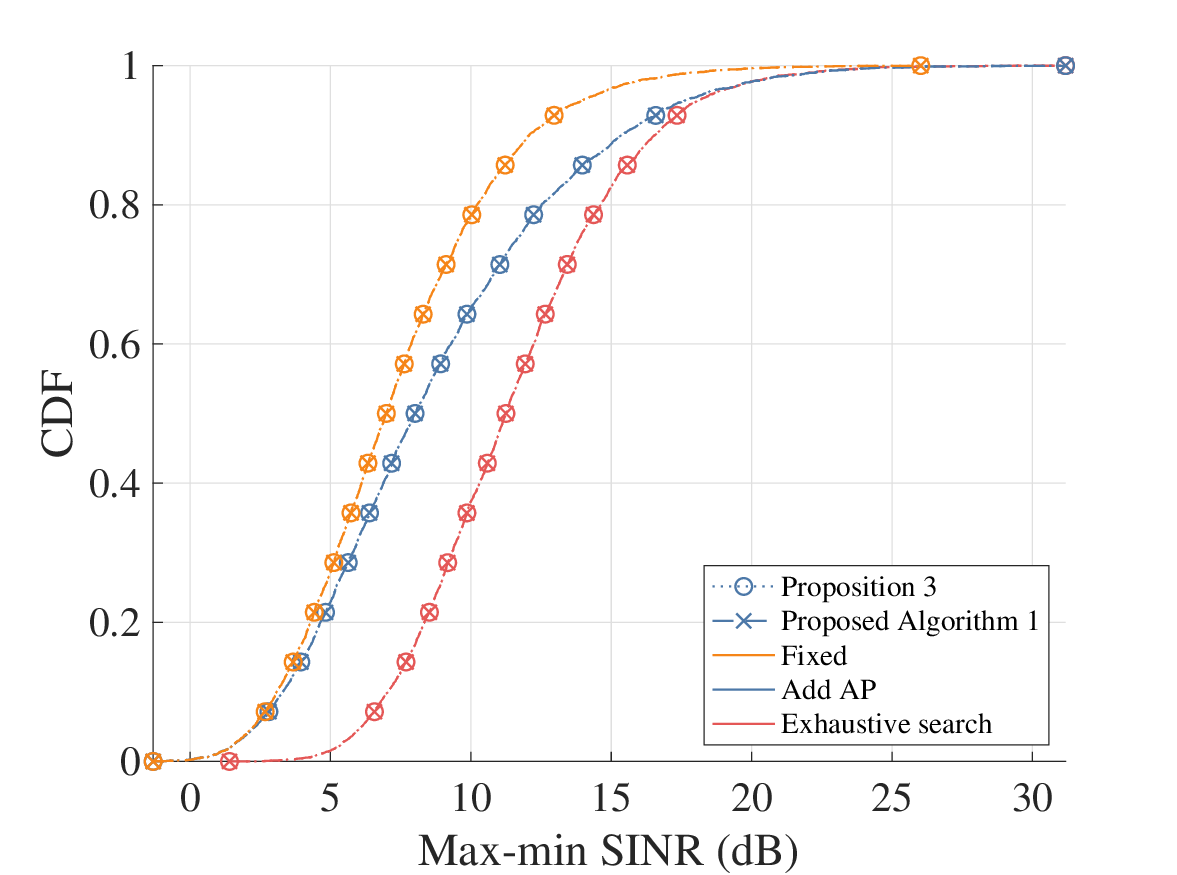}
    \caption{Comparison of Proposition~\ref{proposition_5_2} and proposed Algorithm~\ref{5_algo} with various AP clustering schemes on a simple network setup.}
    \label{fig_5_1}
\end{figure}

Fig.~\ref{fig_5_1} compares the theoretical max-min SINR from \eqref{eq_5_22} in Proposition~\ref{proposition_5_2}, with the converged numerical solution from the proposed Algorithm~\ref{5_algo}. The analysis uses a simple network with $R=9$ APs, $K=4$ antennas per AP, and $N=6$ single-antenna users. To limit the complexity of selecting optimal AP clusters for all users in \eqref{eq_5_23}, a candidate AP set size of $\left|\mathcal{M}_n\right| = 3$ per user is considered, restricting the number of possible AP set combinations. The possible AP sets are determined by the \textit{fixed}, \textit{add AP}, or \textit{exhaustive search} schemes applied to each user's candidate AP set. The maximum transmit power for each user is $\mathrm{p_{max}} = 30\,\mathrm{dBm}$ \cite{Miretti22}. The overlapped cumulative distribution functions (CDFs) of the max-min SINR for varying AP clustering schemes confirm that the algorithm converges to $\mathbf{p}^*$, as per \eqref{eq_5_24} in Proposition~\ref{proposition_5_2}, solving the conditional eigenvalue problem in \eqref{eq_5_21}.

\begin{figure}
    \centering
    \includegraphics[width=\linewidth]{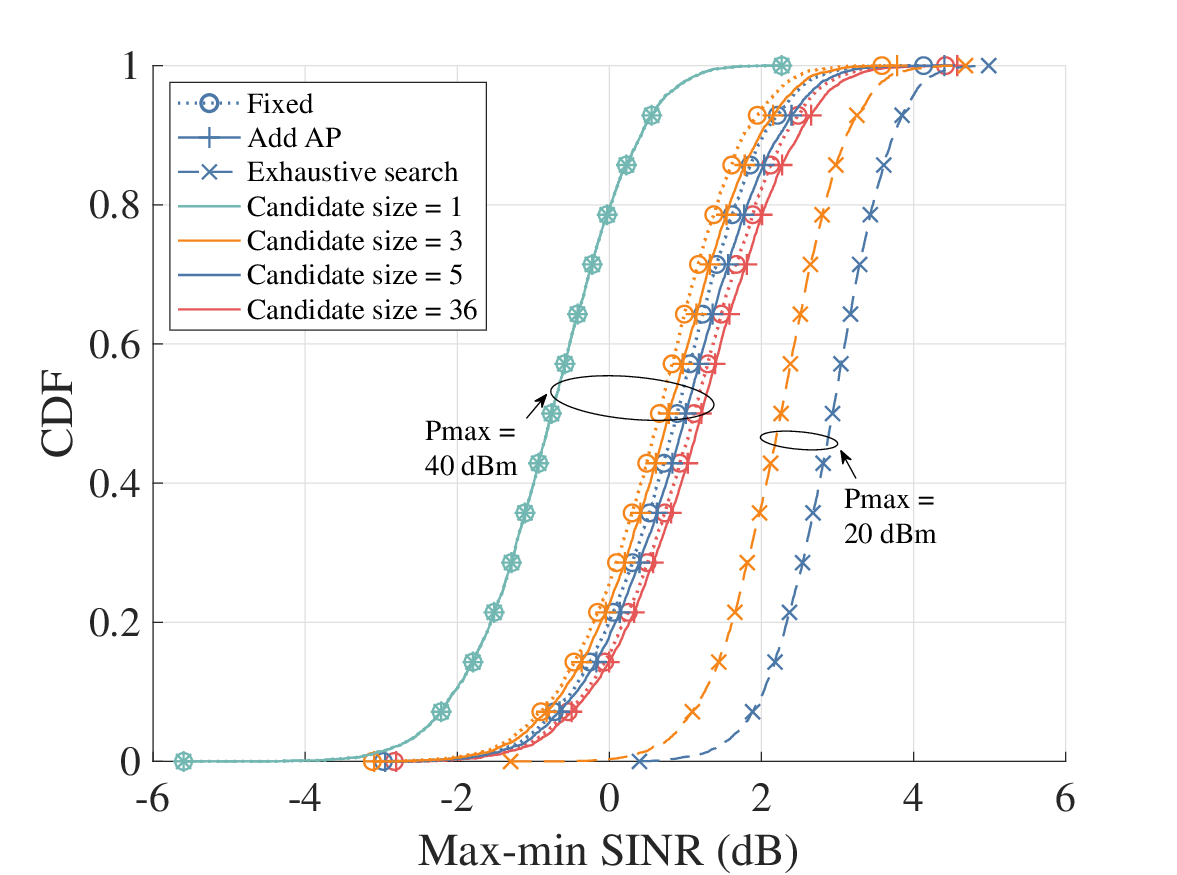}
    \caption{Performance with various AP clustering schemes and candidate set sizes using the proposed Algorithm~\ref{5_algo}. The setup consists of 36 APs, 4 AP antennas, and 40\% user density.}
    \label{fig_5_2}
\end{figure}

In Fig.~\ref{fig_5_2}, we consider a cell-free network with $R=36$ APs, $K=4$ antennas per AP, and $N=58$ users, which is approximately 40\% of the total $RK$ receiver antennas. We refer to this as a 40\% user density. The CDFs of the max-min SINRs are plotted under various clustering schemes and candidate AP set sizes using the proposed Algorithm~\ref{5_algo}. The \textit{exhaustive search} scheme was limited to 5 candidate APs to reduce complexity. The max-min SINR for all schemes increases with the candidate set size due to macro-diversity, with significant gains from size 1 to 3, highlighting the benefits of AP cooperation. However, for the \textit{fixed} and \textit{add AP} schemes, the achievable max-min SINR saturates at lower values, even when the entire network serves and users are allocated a higher power budget. This is due to the weaker signals at distant APs, coupled with poorly managed higher interference at those APs. These schemes are ineffective at mitigating high-interference APs within clusters. Therefore, assigning APs to users based solely on average channel strength is insufficient with MRC receivers. With a limited candidate set size of 3 and a lower user power budget of 20 dBm, the \textit{exhaustive search} scheme achieves higher max-min SINR due to superior interference coordination. The candidate set size is a configurable parameter determined by network operators, and selecting a small candidate set size reduces the computational complexity of the \textit{exhaustive search} scheme in practical implementations.

\begin{figure}
    \centering
    \includegraphics[width=\linewidth]{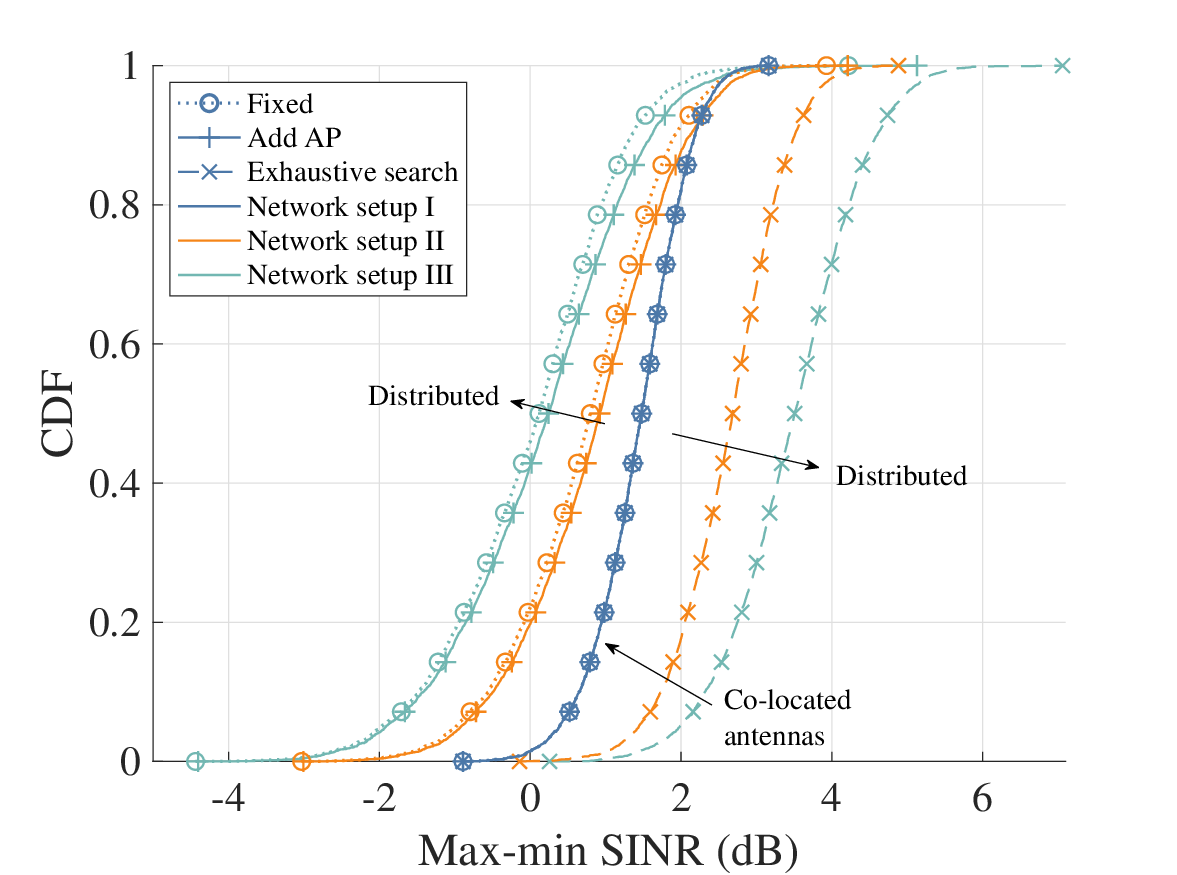}
    \caption{Performance with various AP clustering schemes on different network setups in Table~\ref{table_3_2}, using the proposed Algorithm~\ref{5_algo}. The setups consist of 40\% user density, and $\mathrm{p_{max}} = 20\,\mathrm{dBm}$.}
    \label{fig_5_3}
\end{figure}

In Fig.~\ref{fig_5_3}, we compare the performance of AP clustering schemes across different antenna distributions within the same total geographical area. Table~\ref{table_3_2} outlines the network setups: Setup I represents a co-located antenna deployment, while Setups II and III represent distributed antenna deployments. The setups differ in inter-AP distances and each consists of $N=58$ users, with each user having $\mathrm{p_{max}} = 20\,\mathrm{dBm}$. To ensure a fair comparison, the candidate set size is adjusted so that each user can potentially be served by the same number of antennas, with Setup I restricted to a single AP per user. Accordingly, Setup I corresponds to a cellular network, whereas Setups II and III represent cell-free networks.

\begin{table}[h!]
    \centering
    \caption{\scshape Network Setups}
    \label{table_3_2}
    \begin{tabular}{|l|c|c|c|} \hline 
    &Number of APs &Per AP antennas &Candidate size \\ 
    &($R$) &($K$) &($\left|\mathcal{M}_n\right|$)\\ 
    \hline\hline
        I& 9& 16& 1\\ \hline
        II& 36& 4& 4\\ \hline
        III& 72& 2& 8\\ \hline    
    \end{tabular}
\end{table}

Fig.~\ref{fig_5_3} presents the CDFs of the achievable max-min SINRs under various clustering schemes for the network setups in Table~\ref{table_3_2}, using the proposed Algorithm~\ref{5_algo}. As antennas become more distributed, the performance of the \textit{fixed} and \textit{add AP} schemes degrades due to reduced micro-diversity at each AP, as micro-diversity is crucial for mitigating interference with MRC \cite{Molisch22}. Moreover, these schemes make inefficient use of the macro-diversity available from different APs. Fixed clusters show the sharpest decline in performance due to their inability to avoid interference, which becomes more pronounced in highly distributed setups where interfering users are closer to the APs. Conversely, the \textit{exhaustive search} scheme performs better in distributed scenarios by effectively leveraging the spatial domain by avoiding interference. This behavior contrasts with \cite{Molisch22}, where clustering was not employed. Additionally, distributed cell-free setups provide users with multiple nearby AP connections, improving robustness against shadowing and enhancing fairness.

\section{Conclusion} \label{conclusion}

We re-purposed an iterative power control algorithm based on non-linear Perron-Frobenius theory to jointly optimize uplink power and dynamic user-centric AP clusters, achieving max-min SINR fairness in a cell-free network. Additionally, we introduced a theoretical analysis by framing the joint optimization problem as a conditional eigenvalue problem. The max-min SINR solution is characterized by the spectral radius of a centrally constructed matrix defined by the problem parameters.

Our results show that AP selection and power control are essential for effective interference management with the MRC receiver. With proper coordination, a limited number of APs can outperform the canonical cell-free system using the entire network. We further demonstrated that, with refined clustering, a distributed antenna cell-free setup surpasses the co-located antenna cellular setup by leveraging macro-diversity gains. This enables the MRC receiver, which maximizes desired signal power while ignoring interference, to achieve higher max-min SINRs. Joint optimization with adaptive candidate set sizes in heterogeneous networks is a promising future direction.
\bibliographystyle{IEEEtran}
\bibliography{mybib}

\end{document}